\documentclass[a4paper, 12pt]{article}

\usepackage[dvips]{graphics}
\usepackage[dvips]{epsfig}

\def\s{\scriptscriptstyle}
\def\b{\begin{eqnarray}}
\def\e{\end{eqnarray}}
\def\1{\hskip1pt}
\def\2{\hskip2pt}
\def\4{\hskip4pt}
\def\5{\hskip5pt}
\def\={\5 = \5}
\def\+{\5 + \5}
\def\-{\5 - \5}
\def\k{\kappa}
\def\nn{\nonumber}
\def\df{\dot{f}}
\def\da{\dot{a}}
\def\ddf{\ddot{f}}
\def\dg{\dot{g}}
\def\ddg{\ddot{g}}
\def\dt{\partial^{\phantom{2}}_t}

\renewcommand{\thefootnote}{\fnsymbol{footnote}}

\begin{document}

\begin{center}
{\huge\textbf{Standard Cosmology from Sigma-Model \\}}
\vspace {10mm}
{\large Conall Kennedy\footnote{e-mail: conall@maths.tcd.ie} and 
\setcounter{footnote}{6}
Emil M. Prodanov\footnote{e-mail: prodanov@maths.tcd.ie} \\}
\vspace {6mm}
{\it School of Mathematics, Trinity College, University of Dublin, Ireland} 
\end{center}
\renewcommand{\thefootnote}{\arabic{footnote}}
\setcounter{footnote}{0}
\vspace{10mm}

\begin{abstract}

We investigate (4+1)- and (5+0)-dimensional gravity coupled to a 
non-compact scalar field sigma-model and a perfect fluid within the 
context of the Randall--Sundrum scenario. We find cosmological 
solutions with a rolling fifth radius and a family of warp factors. 
Included in this family are both the original Randall--Sundrum 
solution and the self-tuning solution of Kachru, Schulz and 
Silverstein. Our solutions exhibit conventional cosmology.

\end{abstract}

\normalsize

\newpage

\section{Introduction}

Theories with extra dimensions where our four-dimensional world is a 
hypersurface (three-brane) embedded in a higher-dimensional spacetime and at 
which gravity is localised have been the subject of intense scrutiny since 
the work of Randall and Sundrum \cite{rs}. The main motivation for such models 
comes from string theory where they are reminiscent of the Ho\v{r}ava-Witten 
solution \cite{hw} for the field theory limit of the strongly-coupled $E_8 
\times E_8$ \linebreak heterotic string. The Randall--Sundrum (RS) scenario 
may be modelled \cite{gubser} and \cite{hollo} by coupling gravity to a 
scalar field and mapping to an equivalent supersymmetric quantum mechanics 
problem. A static metric is obtained with a warp factor determined by the 
superpotential. A generalisation to non-static metrics was considered by 
Bin\'{e}truy, Deffayet and Langlois who modelled brane matter as a 
perfect fluid delta-function source in the five-dimensional 
Einstein equations \cite{bdl}. 
However, this resulted in non-standard cosmology in that the square of the 
Hubble constant on the brane was not proportional to the density of the 
fluid. Other cosmological aspects of ``brane-worlds'' have been considered in 
\cite{cosmo}.\\
In this letter we investigate cosmological solutions of five-dimensional 
gravity coupled to a scalar field sigma-model. In much of the current 
literature it is assumed that such scalars depend only on the fifth dimension 
and that the target space metric is of Euclidean signature. By contrast, we 
consider a non-compact sigma-model and allow the scalars 
to depend on time as well as the fifth dimension, which we take to be 
infinite in extent. We also include a perfect fluid with energy-momentum 
tensor $\tilde{T}^\mu_{\phantom{\mu}\nu} \= \mbox{diag\1}(-\rho, p, p, p, P)$ 
and equations of state $P = \omega \rho, \2 p = \tilde{\omega} \rho$. A 
family of warp factors that includes both the original RS solution and the 
self-tuning solution of Kachru, Schulz and Silverstein \cite{kachru} is 
found. The fifth radius is time-dependent. We find that the fluid exists 
provided $\omega = \tilde{\omega} = 1$. Conventional cosmology is also 
obtained.\\
It may appear somewhat unnatural to have an indefinite target space metric 
since some of the scalars then have ``wrongly-signed'' kinetic terms.
However, such scalars have been considered before in the literature. Within 
the context of $d+1$ gravity they are descended from vector fields after 
dimensional reduction along a timelike direction of a higher dimensional 
``two-time'' theory \cite{hull} and \cite{stelle}, whilst in $d+0$ dimensions 
they are interpreted as axions after dualisation of a ($d-$1)-form field 
strength \cite{crem1}, \cite{crem2} and \cite{gid}. Thus, our paper should be 
interpreted in the light of these works. 

\section{The Model}

We shall present our calculations in $(4+1)$-dimensional spacetime and only 
quote analogous results for the $5+0$ case. The action for gravity coupled to 
two scalars is:
\b
S \ = \int d^{\s 4}x \2 dr \5 (\mathcal{L}^{\s (5)}_{\s MATTER} \+ 
\mathcal{L}^{\s (5)}_{\s GRAVITY}) \2,
\e
where:
\b
\mathcal{L}^{\s (5)}_{\s MATTER} & = &\!\!\!\! - 
\frac{1}{2} \sqrt{\!-g^{\s (5)}} 
\nabla^{\mu} \phi_{i} \nabla^{\nu} \phi_{j} G^{ij}(\phi) g^{\s (5)}_{\mu\nu} -
\!\sqrt{\!-g^{\s (5)}} U(\phi) - \!\sqrt{\!-g^{\s (4)}} V(\phi) \delta(r)\2, 
\nn \\
\mathcal{L}^{\s (5)}_{\s GRAVITY}\!\! & = & \!\!\frac{1}{\k^2} \2 
\sqrt{-g^{\s (5)}} \2\1 R.
\e
Here, $g^{\s (4)}_{\mu\nu}$ is the pull-back of the five-dimensional metric 
$g^{\s (5)}_{\mu\nu}$ to the (thin) domain wall taken to be at $r=0$. The 
wall is represented by a delta function source with coefficient 
$V(\phi)$ parametrising its tension. 
We take $G_{ij} = \mbox{diag\1}(1,-1)$. The ``correctly-signed'' scalar, 
$\phi^1$, may be interpreted as the dilaton and the ``wrongly-signed'' 
scalar, $\phi^2$, as an axion. (It is possible to consider a non-trivial 
coupling between the two --- for example, 
$G_{ij} = \mbox{diag\1}(1,-e^{\sigma \phi^1})$ is discussed in \cite{crem2}.) 
\\
We assume a separable metric with flat spatial three-sections on the wall:
\b
\label{metric}
ds^2 = - e^{-A(r)} dt^2 + e^{-A(r)}g(t)(dx^2 + dy^2 + dz^2) + f(t)dr^2. 
\e
This is a natural generalisation of the 4$d$ flat Robertson-Walker metric to 
a RS context.\\ 
Given the above ansatz, it is not unreasonable to assume scalars of 
the form
\b
\label{dilaton}
\phi^i (t, r) \= a^i \2 \psi(t) \+ b^i \2 \chi(r) \2.
\e
Since $\phi^i$ can be considered as coordinates on the target spacetime we 
must require them to be linearly independent. This imposes the condition
\b
\mbox{det\1}\left(\matrix{a^1 & b^1 \cr a^2 & b^2}\right) \ne 0 \2.
\e
The Schwarz inequality $\frac{(a\cdot a)(b\cdot b)}{(a\cdot b)^2}\2 < 1$ follows as a corollary.\\
We also make the ansatz that both the potentials $U$ and $V$ are of Liouville 
type (see, for instance, \cite{reall}):
\b
\label{pot} 
V(\phi) & = & V_0 e^{\alpha_i \phi^i} \2, \nn \\
U(\phi) & = & U_0 e^{\beta_i \phi^i} \2.
\e
The energy--momentum tensor for the scalar fields is:
\b
T^{\s (0)}_{\mu\nu} & = & \frac{1}{2} \2 \nabla_{\mu} \phi^{i} \2 \nabla_{\nu} 
\phi^{j} \2 G_{ij} - \frac{1}{2} \2 g_{\mu\nu}\Bigl(\frac{1}{2} \2 
\nabla_{\alpha} \phi^{i} \2 \nabla_{\beta} \phi^{j} \2 G_{ij} \2 
g^{\alpha\beta} + U(\phi)\Bigr) \nn \\ \nn \\&& \hskip120pt - \2\frac{1}{2} \2 \frac{\sqrt{-g^{\s (4)}}}
{\sqrt{-g^{\s (5)}}} 
\2 V(\phi) \2 \delta(r) \2 g^{\s (4)}_{ab} \2 \delta^a_\mu \2 \delta^b_\nu \2.
\e
We introduce a perfect fluid via its energy--momentum tensor:
\b
\tilde{T}^\mu_{\phantom{\mu}\nu} \= \mbox{diag\1}(-\rho, p, p, p, P) 
\e
with $\rho$ the density and $p$ and $P$ the pressures in the $x, y, z$ and 
fifth dimensions respectively. The preferred coordinate system (\ref{metric}) 
is taken as the rest frame of the fluid. \\
Einstein's equations \5 
$ G_{\mu\nu} \= \k^2 (T^{\s (0)}_{\mu\nu} + \tilde{T}_{\mu\nu})$ \5 reduce to:
\b
\label{1} 
&&\frac{1}{4} \1 \frac{\df}{f} \1 \frac{\dg}{g} + \frac{\dg^2}{g^2} +
\frac{1}{4} \1 \frac{\df^2}{f^2} - \frac{1}{2} \1 \frac{\ddf}{f}  - 
\frac{\ddg}{g} - \frac{\k^2}{2}\1 a\cdot a \1 \dot{\psi}^2 - \k^2 \1 
e^{-A} (\rho + p) = 0 \1, \\ \nn \\
\label{4} 
&&\frac{3}{4} \2 \frac{\df}{f} \2 \frac{\dg}{g} \+ \frac{3}{4} \2 
\frac{\dg^2}{g^2} \- \frac{\k^2}{4} \2 a\cdot a \2 \dot{\psi}^2 \- 
\k^2 \2 e^{-A}\rho \=  0 \2, \\ \nn \\
\label{5} 
&&\frac{3}{2} \2 (A^{\prime 2} - A^{\prime\prime}) \+ \frac{\k^2}{4} \2 
b\cdot b \2 \chi^{\prime 2} \+ \frac{\k^2}{2} \2 f \2 U  \+  \frac{\k^2}{2} 
\2  f^{1/2} \2 V \2 \delta(r) \1 = \1 0 \1, \\ \nn \\
\label{6} 
&& \frac{3}{2} \2 \frac{\ddg}{g} \+ \frac{\k^2}{4} \2 a\cdot a \2 \dot{\psi}^2
\+ \k^2 \2 e^{-A} \2 P \= 0 \2, \\ \nn \\
\label{8} 
&& \frac{3}{2} \2 A^{\prime 2} \- \frac{\k^2}{4} b\cdot b \2 \chi^{\prime 2} \+
\frac{\k^2}{2} \2 f \2 U  \= 0 \2, \\ \nn \\
\label{9} 
&& \frac{3}{2} \2 A^\prime \2 \frac{\df}{f} \+ \k^2 \2 a\cdot b \2
\dot{\psi}\2 \chi^{\prime} \= 0 \2. 
\e
In the above equations we have assumed separability. This requires that the 
density and pressures are each of the form $e^{A(r)}$ times a function of 
$t$. We are interested in solutions with $\df \neq 0$. This requires 
$a\cdot b \neq 0$, as can be deduced from (\ref{9}).\\ 
The equations of motion for the scalar fields
\b
\label{eqm}
\nabla^2 \phi^j G_{jk} \- \frac{\partial U(\phi)}{\partial \phi^k} \-
\frac{\sqrt{-g^{\s (4)}}}{\sqrt{-g^{\s (5)}}} \2 \frac{\partial V(\phi)}
{\partial \phi^k} \2 \delta(r) \= 0
\e
result in the following bulk equations
\b
\label{12}
\dt(f^{1/2}g^{3/2} \dot{\psi}) & = & 0, \\
\label{bulk}
b_i \2 (2 A^\prime \chi^\prime - \chi^{\prime\prime}) \+ f \2 \beta_i \2 U_0 
& = & 0,
\e
and the jump condition:
\b
\label{jump}
\lim_{\s \epsilon \to 0^{\s +}} \2 \biggl[ b_i 
\Bigl(\chi^\prime(\epsilon) - \chi^\prime(-\epsilon)\Bigr)\biggr] \= \alpha_i \2 f^{1/2} \2 V(\phi(t,0)).
\e

\section{The Solutions}
 
Equation (\ref{9}) implies that we can make 
the following choice:
\b
\label{scalar1}
\k \chi^\prime(r) & = & \sqrt{6} \2 A^\prime(r) \2,  \\
\label{scalar2}
\k \dot{\psi}(t)  & = & -\frac{\sqrt{6}}{4} \2 \frac{1}{a\cdot b} \2 
\frac{\df(t)}{f(t)} \2.
\e
\vskip5pt
\noindent
\underline{\it{The Warp Factor}}
\vskip5pt
\noindent
Inserting (\ref{scalar1}) into (\ref{8}) gives $U(\phi)$ as:
\b
\label{U}
U \= - \frac{3}{\k^2} \2 \frac{1}{f} \2 A^{\prime 2} \2 (1 - b\cdot b) \2.
\e
We can express the domain wall potential \5 $V(\phi) \2 \delta(r)$ \5 as \5
$V(\phi)\delta(r) \= \linebreak V_0 \1 f(t)^{-1/2} \1 \delta(r)$. 
Equation (\ref{5}) can then be rewritten in the form
\b
\label{rdep}
A^{\prime \prime} - 2 b\cdot b A^{\prime 2} - \frac{\k^2}{3}\1 V_0 \1 
\delta(r) \1 = \1 0,
\e
yielding the following options for $A(r)$ and $V_0$:
\vskip8pt
{\bf 1.} If $b\cdot b = 0$, we find $A(r) = 2 \sigma k|r|$, where $\sigma = 
\pm 1$. Then $V_0 = 12 \sigma k \k^{-2}$. $\sigma = -1$ is the RS1 solution 
and $\sigma = +1$ is the RS2 solution, as described in \cite{csaki}.
\vskip8pt
{\bf 2.} If $b\cdot b \ne 0$, we find  $A(r) = \xi \ln(k|r| + 1)$ where
$\xi = -\frac{1}{2 b\cdot b}$ and $V_0 = -\frac{3 k \k^2}{b\cdot b}\2.$ 
If $b\cdot  b$ and $k$ are both positive, then this represents the self-tuning 
solution of Kachru, Schulz and Silverstein \cite{kachru}. As observed in 
\cite{youm} and \cite{gomez}, if $k<0$ there are naked singularities at 
$|r| = -1/k$ whose interpretation is currently of some debate \cite{kak}. 
\vskip8pt
\noindent
The above forms for $U$ and $V$ are consistent with (\ref{pot}) if $\alpha_i =
 \frac{\beta_i}{2} = \frac{2 \k b_i}{\sqrt{6}}$ and $U_0 = - \frac{3}{\k^2} \2 
A^{\prime 2}(0) \2 (1 - b\cdot b)$. It can now be verified that
 (\ref{bulk}) is equivalent to (\ref{rdep}) in the 
bulk, whilst (\ref{jump}) yields no further information.

\vskip8pt
\noindent
\underline{\it {The Cosmology}}
\vskip5pt
\noindent The equation of motion (\ref{12}) implies that
\b
\label{psi}
\dot{\psi}(t) \= \frac{1}{\k} \2 f(t)^{-1/2} \2 g(t)^{-3/2}\2.
\e
This assumes $f$ is not constant, otherwise (\ref{12}) is trivially satisfied 
due to (\ref{scalar2}).
We find that $f(t)$ and $g(t)$ are related via the 
following equation:
\b
\label{fg}
\frac{\df(t)}{f(t)^{1/2}} = \mu \2 g(t)^{-3/2} \2,
\e
where $\mu \= - \frac{4a\cdot b}{\sqrt{6}}.$ \\
Adding equations (\ref{4}) and (\ref{6}) gives:
\b
\label{c1}
\dg^2 \+ 2g\ddg \+ \frac{\df}{f} \2 \dg g \+ \frac{4}{3} \2 \k^2 \2
g^2 \2 e^{-A} \2 (P \-  \rho) \= 0.
\e
On the otherhand, using (\ref{4}) and (\ref{12}) in (\ref{1}) we obtain:
\b
\label{c2}
\dg^2 \+ 2g\ddg \+ \frac{\df}{f} \2 \dg g  \+ 2 \k^2 g^2 \2 e^{-A} \2 
(p \- \rho) \= 0.
\e
Consequently, the relation
\b
\label{ttwiddle}
p \= \frac{1}{3} \2 \rho \+ \frac{2}{3} \2 P \2,
\e
may be deduced. \\
We now assume the equation of state $P = \omega \rho$ or, equivalently, 
$p = \frac{1}{3}(1+2\omega)\rho \equiv \tilde{\omega} \rho$.
From (\ref{4}) and (\ref{scalar2}), the density $\rho$ is given by
\b
\label{dense}
\rho(t, r) \= \frac{3e^A}{4\k^2}\biggr( \frac{\df}{f} \2 \frac{\dg}{g} \+ 
\frac{\dg^2}{g^2}\2 - \frac{a\cdot a}{8(a\cdot b)^2}\2 
\frac{\df^2}{f^2}\biggl), 
\e
so that (\ref{c1}) may be alternatively expressed as:
\b 
\label{g2}
 \omega \2 \frac{\dg^2}{g^2} \+ 2\frac{\ddg}{g} \+ \omega \2 
\frac{\df}{f}\2 \frac{\dg}{g} \2 \+ 
(1-\omega) \2 \frac{a\cdot a}{8(a\cdot b)^2} \2 \frac{\df^2}{f^2} \= 0. 
\e
Taken together with (\ref{fg}), equation (\ref{g2}) defines the cosmology. \\
We seek either power law, $f \sim t^q$, or exponential (inflationary), $f 
\sim e^{\gamma t}$, solutions of (\ref{g2}). The corresponding solutions for 
$g(t)$ are $g \sim t^{(2-q)/3}$ and $g \sim e^{-\gamma t/3}$ 
respectively. The exponents $q$ and $\gamma$ are non-zero but otherwise 
arbitrary. \\
There are two cases to consider: $\omega = 1 \mbox{ and } \omega \neq 1$. 
\vskip2pt
\noindent
\underline{(A) $\omega = 1$}\\
From (\ref{dense}), it follows that in the exponential case the density 
is positive provided 
\5 $\frac{a\cdot a}{(a\cdot b)^2} < -\frac{16}{9}$ 
(independently of $\gamma$). On the otherhand, the density is positive in the 
power law case provided 
\5 $\frac{a\cdot a}{(a\cdot b)^2} < h(q) \equiv \frac{16}{9}\2 
\frac{(2-q)(1+q)}{q^2}$. 
As shown in the figure below,
\begin{figure}[!h]
\begin{center}
\mbox{\epsfxsize 90mm
  \epsfbox{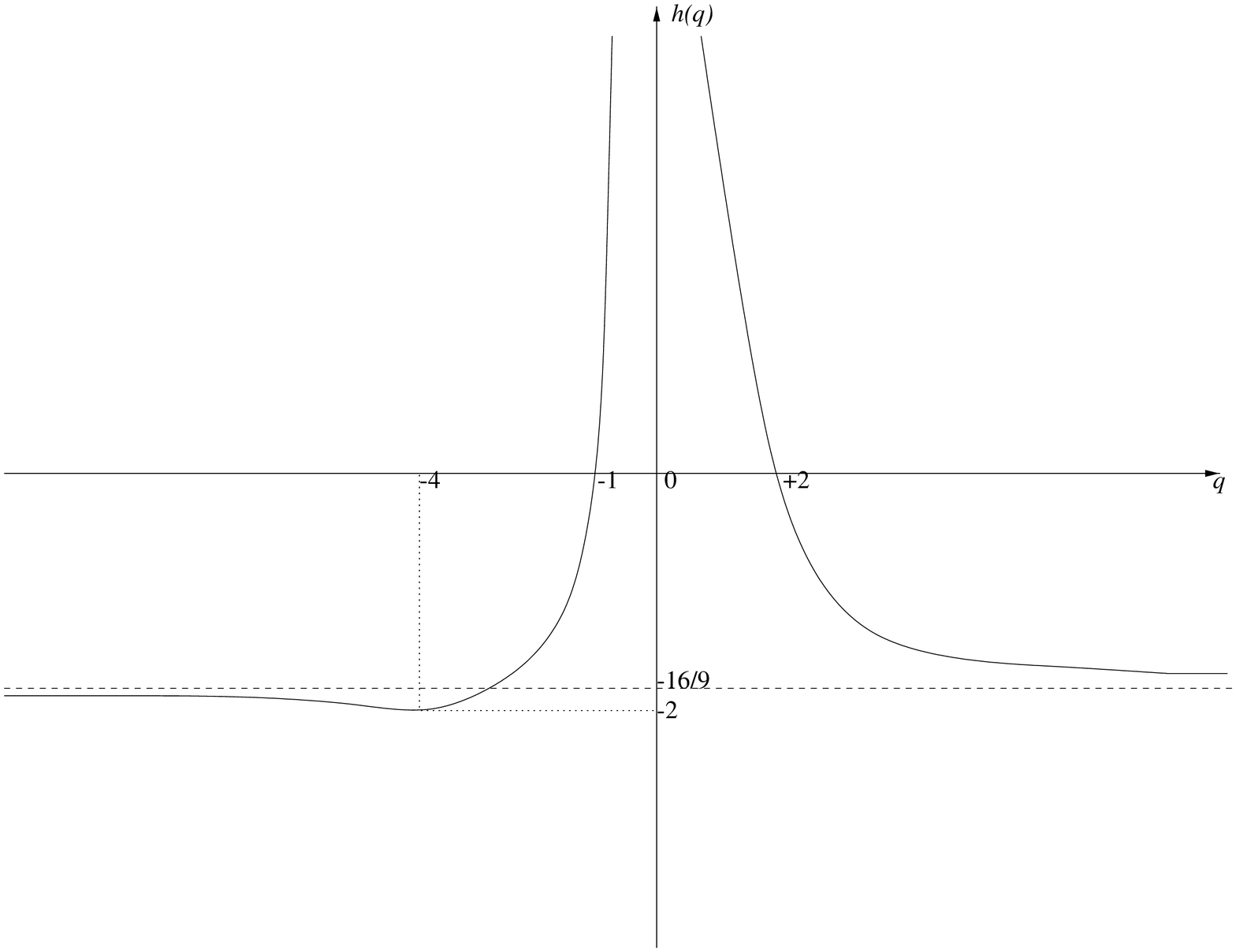}}
\parbox{90mm}{\scriptsize Figure: A sketch of the function $h(q) = 
\frac{16}{9}\2 \frac{(2-q)(1+q)}{q^2}$. The function has a minimum of $-2$ at 
$q=-4$ and tends to $-\frac{16}{9}$ as $q \rightarrow \pm \infty$.}
\end{center}  
\end{figure} 
 the minimum of $h(q)$ is $-2$ so 
we can achieve positive density for all $\gamma$ and $q$ if we choose 
\5 $\frac{a\cdot a}{(a\cdot b)^2} < -2^{}$\1\footnote{This choice is 
consistent with the Schwarz inequality provided $b\cdot b > -\frac{1}{2}$. 
If $b\cdot b \leq -\frac{1}{2}$, positive density is achieved only for a 
limited range of $q$.}. 
Defining the scale factor and Hubble constant as per usual by $a^2(t) = g(t)$ 
and $H = \da/a$, it is easy to see that we obtain conventional cosmology, 
$H^2 \propto \rho$, for both the power law and exponential cases.
\vskip2pt
\noindent
\underline{(B) $\omega \neq 1$}\\
Solution of (\ref{g2}) leads to the above inequalities for 
$\frac{a\cdot a}{(a\cdot b)^2}$ becoming strict equalities which, in turn, 
leads to the vanishing of the density. Hence, the fluid only exists if 
$\omega = 1$.

\vskip8pt
\noindent
\underline{\it{The Euclidean Case}}
\vskip5pt
\noindent The only essential difference between the 5+0 case and the 4+1 case 
considered above is that $\tilde{T}^\mu_{\phantom{\mu}\nu}$ flips sign. This 
changes the sign of $\rho$ in (\ref{dense}) so that the density is positive if 
\5 $\frac{a\cdot a}{(a\cdot b)^2} > -2^{}$. Similar considerations (see 
previous footnote) apply as to the range of $q$.

\section{Discussion}

We note in passing that the scalar field equations of motion, (\ref{eqm}), 
imply that $\nabla^\mu \2 T^{\s (0)}_{\mu\nu} = 0$ (and conversely 
off the brane only). This, in turn, implies that the fluid equation of 
motion $\nabla_\mu \2 \tilde{T}^\mu_{\phantom{\mu}\nu} = 0$ is automatically 
satisfied. In this sense, the same results in the bulk can be obtained from 
Einstein's equations and 
$\nabla_\mu \2 \tilde{T}^\mu_{\phantom{\mu}\nu} = 0$. \\
It may seem a bit unusual to consider a non-static fifth radius (some 
authors \cite{f-const} give arguments against rolling dilatons). We would like 
to present an intuitive argument in favour of our choice. Consider a 
five-dimensional spacetime with Robertson--Walker metric:
\b
\label{one}
ds^2 \= -dt^2 \+ g(t)\Bigl(dx^2 + dy^2 + dz^2 + dR^2\Bigr) \2.
\e
The $(x, y, z, R)$-space is isotropic. Change coordinates via
\b
dr \= e^{-\frac{1}{2}A(r)} \2 dR \2
\e
and perform a conformal transformation of the metric:
\b
ds^2 \to e^{-A(r)} \2 ds^2.
\e 
Then the metric becomes:
\b
ds^2 = - e^{-A(r)} dt^2 + e^{-A(r)}g(t)(dx^2 + dy^2 + dz^2) + g(t) dr^2
\e
\noindent 
$\!\!$The warp factor of the conformal transformation violates the symmetry 
between the four spatial coordinates. Zel'dovich \cite{zel} gives arguments 
that any universe will become isotropic with time and non-isotropic expansion 
causes particle creation. To avoid particle creation in the bulk one could 
restore isotropy by ``untwisting'' the fifth dimension with another warp 
factor, i.e., replacing $g(t)$ by another function of time, $f(t)$, such that 
the four spatial dimensions are still isotropic. \\
Within our model we can still have scalar fields depending on brane 
coordinates if we require a static fifth radius. In this case we need to 
introduce viscosity into the fluid by making 
$\tilde{T}^\mu_{\phantom{\mu}\nu}$ non-diagonal\footnote{We are grateful to 
Brian Dolan for discussions on this point.}. The sum of the energy-momentum 
tensors of the scalar fields and the fluid should then amount to a purely 
diagonal tensor. \\
From our initial separability assumptions and from equation (\ref{dense}) it 
is clear that if the warp factor decreases with $r$ then the density grows 
without limit as we go off the brane and the fluid is smoothly distributed 
over the entire extra dimension. \\
Considering a thick brane (in Lorentzian or Riemannian signature) 
within our model is straightforward. Thickening the brane requires only 
smearing the delta function in the domain wall potential by expressing it as a
limit of some delta-sequence, for example, $\delta_\nu(r) = \frac{1}{\pi}
\frac{\nu}{1+\nu^2 r^2}$ \1, where $\frac{1}{\nu}$ parametrises the brane 
thickness. \\
From (\ref{1}) -- (\ref{9}) it is evident that under the transformation 
$f \to -f$ the potentials $U$ and $V$ change sign but otherwise the analysis 
is unmodified. Thus one can make the fifth dimension timelike rather than 
spacelike. Such a possibility was alluded to in \cite{over} and 
\cite{Chaich}.\\
Finally, it would be interesting to see if our model(s) can be embedded in 
five-dimensional Lorentzian or Euclidean supergravity, as has recently been 
done for the minimal Randall--Sundrum model in 4+1 dimensions 
\cite{bag}, \cite{pom}.

\section*{Acknowledgements}

We are sincerely grateful to Siddhartha Sen for a suggestion that initiated 
these investigations and for useful comments. We have benefited from 
fruitful discussions with Brian Dolan, Petros Florides, David Simms,
Charles Nash and Andy Wilkins. C. K. acknowledges the support of Trinity 
College, Dublin and Enterprise Ireland.

\end{document}